\documentclass[aps,prd,twocolumn,groupedaddress,preprintnumbers,
              floatfix, nofootinbib,longbibliography]{revtex4}
            
\usepackage[utf8]{inputenc}

\usepackage{hyperref}
\usepackage{slashed}
\usepackage{natbib}
\usepackage{graphicx}
\usepackage{epsfig}
\usepackage{amsmath}
\input{epsf}
\usepackage{psfrag}

\usepackage[percent]{overpic}

\usepackage[usenames,dvipsnames]{color}

\newcommand{\beq}{\begin{eqnarray}}
\newcommand{\eeq}{\end{eqnarray}}

\newcommand{\nn}{\nonumber \\}

\usepackage{amsmath,color}
\usepackage{amssymb}
\usepackage{bm}
\usepackage{graphicx}
\usepackage{multirow}

\usepackage{soul} 
\newcounter{RSQ}

\begin{document}

\title{Sullivan process near threshold and the pion gravitational form factors  }

\author{Yoshitaka Hatta and Jakob Schoenleber }
\affiliation{Physics Department, Brookhaven National Laboratory, Upton, NY 11973, USA}
\affiliation{RIKEN BNL Research Center, Brookhaven National Laboratory, Upton, NY 11973, USA}

\begin{abstract}

We propose a novel method to experimentally access the gravitational form factors (GFFs) of the charged pion $\pi^+$ through the Sullivan process in electron-proton scattering. We  demonstrate that the cross sections of  $J/\psi$-photoproduction and $\phi$-electroproduction  near the respective thresholds are dominated by the  gluon  GFF of the pion to next-to-leading order in perturbative QCD. We predict cross sections for the Electron-Ion Collider and the Jefferson Lab experiments.  

\end{abstract}

\maketitle

{\it Introduction}---Recently, exclusive photoproduction of heavy vector mesons such as $J/\psi$ and $\Upsilon$ near threshold has emerged as a promising tool to access the gluon gravitational form factors (GFFs) of the nucleon defined as the off-forward matrix elements of the QCD energy momentum tensor $\langle p'|T^{\mu\nu}|p\rangle$ \cite{Hatta:2018ina,Hatta:2019lxo,Mamo:2019mka,Wang:2019mza,Boussarie:2020vmu,Kharzeev:2021qkd,Hatta:2021can,Guo:2021ibg,Mamo:2021krl,Sun:2021pyw,Guo:2023qgu,Wang:2024aqd,Pentchev:2024sho,Hatta:2025vhs,Guo:2025jiz} \cite{GlueX:2019mkq,Duran:2022xag,GlueX:2023pev,Klest:2025rek}.\footnote{See  \cite{Du:2020bqj,JointPhysicsAnalysisCenter:2023qgg,Sakinah:2024cza,Tang:2024pky} for other approaches to $J/\psi$ photoproduction near threshold.} 
This is complementary to the `canonical' approach to extract the  quark GFFs from the generalized parton distributions (GPDs) in deeply virtual Compton scattering (DVCS) \cite{Anikin:2007yh,Burkert:2018bqq,Kumericki:2019ddg,Dutrieux:2024bgc}. The major challenge of GPD-based approaches is to isolate  the contribution of the spin-2 energy momentum tensor from infinitely many leading twist operators with different spins. This problem can be mitigated in near-threshold productions characterized by large values of   the skewness variable  $\xi\sim {\cal O}(1)$. In this kinematical region, the sensitivity to the spin-2 component is significantly enhanced \cite{Hatta:2021can,Guo:2021ibg,Guo:2023qgu} compared to the situation in high energy processes with $\xi\ll 1$. This paves the way to efficiently access the nucleon's gluon GFFs.

Importantly, one can realize the situation $\xi\sim {\cal O}(1)$ in the electroproduction of light vector mesons near threshold where the photon virtuality $Q^2$ serves as the hard scale  for GPD factorization \cite{Hatta:2021can}. The case with $\phi$-production off the nucleon is studied in a separate publication  \cite{Hatta:2025vhs},  where it is demonstrated to next-to-leading order (NLO) in perturbative QCD that the cross section is dominated by the gluon and strangeness GFFs. 

In this paper, for the first time we explore the physics of near-threshold vector meson productions off the charged pion $\pi^+$, with the goal of constraining the pion GFFs. 
The GFFs of the pion are of special  interest due to the unique nature of the pion as the Nambu-Goldstone bosons of  spontaneously broken chiral symmetry. The soft pion theorem imposes various constraints on the GFFs and even GPDs \cite{Novikov:1980fa,Polyakov:1998ze,Chavez:2021llq} which could be experimentally tested. While there has been a recent surge of interest in computing the pion GFFs in various approaches  
\cite{Kumano:2017lhr,Tanaka:2018wea,Tong:2021ctu,Hackett:2023nkr,Xu:2023izo,Wang:2024lrm,Fujii:2024rqd,Broniowski:2024oyk,Liu:2024jno,Cao:2024zlf},  attempts to extract the pion GFFs from experimental data are quite limited   \cite{Kumano:2017lhr}. 
We shall demonstrate that the new observables considered here have very good  sensitivities to  the pion GFFs, and make predictions for future experiments at Jefferson Laboratory (JLab) \cite{JeffersonLabSoLID:2022iod} and the Electron-Ion Collider (EIC) at Brookhaven National Laboratory \cite{AbdulKhalek:2021gbh}.

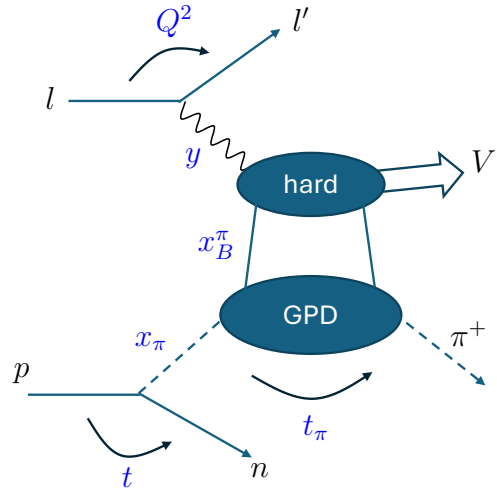
\begin{figure}[b]
\vspace{-5mm}
        \begin{overpic}[width=0.38\textwidth]{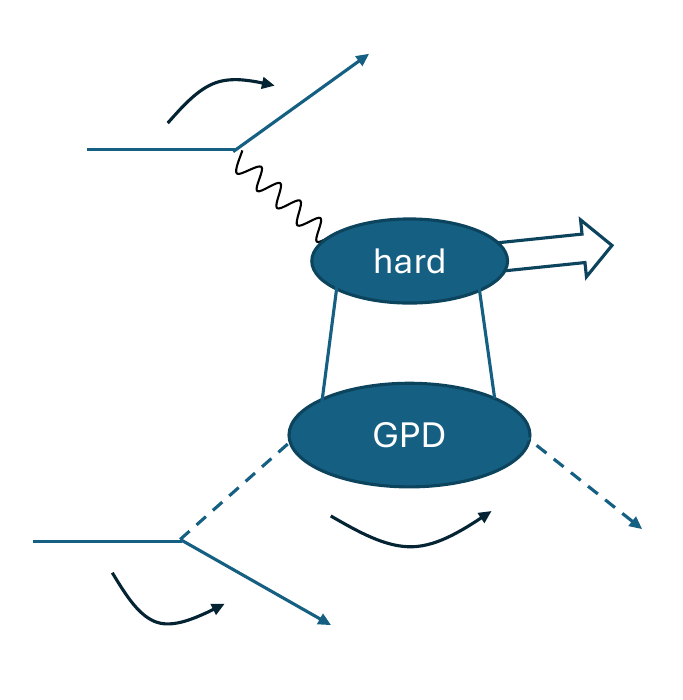}
         \put(29,92){\large{${\color{blue}Q^2}$}}
         \put(55,92){\large{$l'$}}
         \put(8,77){\large{$l$}}
         \put(33,67){ \large{${\color{blue}y}$}}
          \put(89,65){\large{$V$}}
          \put(85,32){\large{$\pi^+$}}
           \put(25,32){\large{${\color{blue}x_\pi}$}}
            \put(2,26){\large{$p$}}
             \put(47,7){\large{$n$}}
              \put(57,15){\large{${\color{blue}t_\pi}$}}
            \put(21,5){ \large{${\color{blue} t}$}}
             \put(37,50){\large{${\color{blue}x_B^\pi}$}}
        \end{overpic}
\vspace{-5mm}
    \caption[*]{ Vector meson production in Sullivan process.}
    \label{dvmp}
\end{figure}

{\it Exclusive vector meson production in Sullivan process}---The Sullivan process  \cite{Sullivan:1971kd} collectively refers to a subset of electron-proton $e+p$ scattering events  where the incoming proton transitions into a neutron by emitting  a virtual pion $p(p)\to \pi^{+*}(p_{\pi}) +n(p_n)$ with virtuality $p_\pi^2=(p-p_n)^2=t<0$, see Fig.~\ref{dvmp}. The pion then interacts with a virtual photon emitted from the electron $e(l)\to \gamma^*(q)+e(l')$ with virtuality $q^2=(l-l')^2=-Q^2\le 0$.   Depending on subsequent $\gamma^*+\pi^{+*}$ reactions, different aspects of the pion structure can be studied \cite{Aguilar:2019teb}.  The reaction $\gamma^* +\pi^{+*}\to \pi^+$ studied in the original paper \cite{Sullivan:1971kd} has become  a cornerstone of modern experimental determinations of the pion electromagnetic form factor    \cite{JeffersonLabFpi-2:2006ysh,JeffersonLab:2008jve}. In the context of the pion GPDs, previous works focused on  Deeply Virtual Compton Scattering (DVCS)  $\gamma^*(q)+\pi^{+*}(p_{\pi}) \to \gamma (q') + \pi^+(p_{\pi}')$ at high center-of-mass energy $s_\pi =(p_\pi+q)^2\sim Q^2$ \cite{Amrath:2008vx,Chavez:2021koz}. There one encounters the usual complications such as the interference with the Bethe-Heilter process and the deconvolution problem of GPDs. In this work, we instead study two exclusive processes: (i) Deeply Virtual Meson Production (DVMP)   $\gamma^*(q)+\pi^{+*}(p_{\pi}) \to V(q') + \pi^+(p_{\pi}')$ where $Q^2\gg \Lambda_{\rm QCD}^2$ and $V$ is a neutral, longitudinally polarized light vector meson such as $\rho^0(770)$, $\omega(783)$ or $\phi(1020)$. (ii) Photoproduction ($Q^2=0$) of  $ J/\psi(3097)$. 
The necessary hard scale  for GPD factorization is provided by $Q^2$ in the former and the $J/\psi$ mass  in the latter. The novelty of our work is that we analyze these processes in the threshold region 
\beq
s_\pi \gtrsim s_\pi^{th}\equiv (m_\pi+m_V)^2.
\eeq  
 (See \cite{Goncalves:2015mbf,Kumar:2022nqu} for  studies in the high energy limit.) 
Some caveats regarding the application of GPD factorization theorems to the threshold region have been discussed in \cite{Hatta:2025vhs}, see also \cite{Hatta:2021can,Guo:2021ibg}.  In particular, $\xi$ cannot be too close to 1, and the preferred region is $\xi= 0.4\sim 0.6$ for the nucleon target. For the pion target, these caveats are less significant due to the light pion mass,  and in fact even a lower region in $\xi$ can be also studied as we shall see. A more serious problem,  which already arises in the context of form factor extractions (see e.g., \cite{JeffersonLab:2008jve,Leao:2024agy}),  may be the fact that the incoming pion is virtual. 
However, as long as $p_\pi^2=t$ is small, the impact on the pion GPDs is expected to be minor \cite{Aguilar:2019teb} and can be theoretically taken into account  \cite{Broniowski:2022iip}.  Assuming this to be the case, our goal is to assess to what extent  the scattering amplitudes  are sensitive to the pion GFFs.

\begin{figure}[t]
        \begin{overpic}[width=0.4\textwidth]{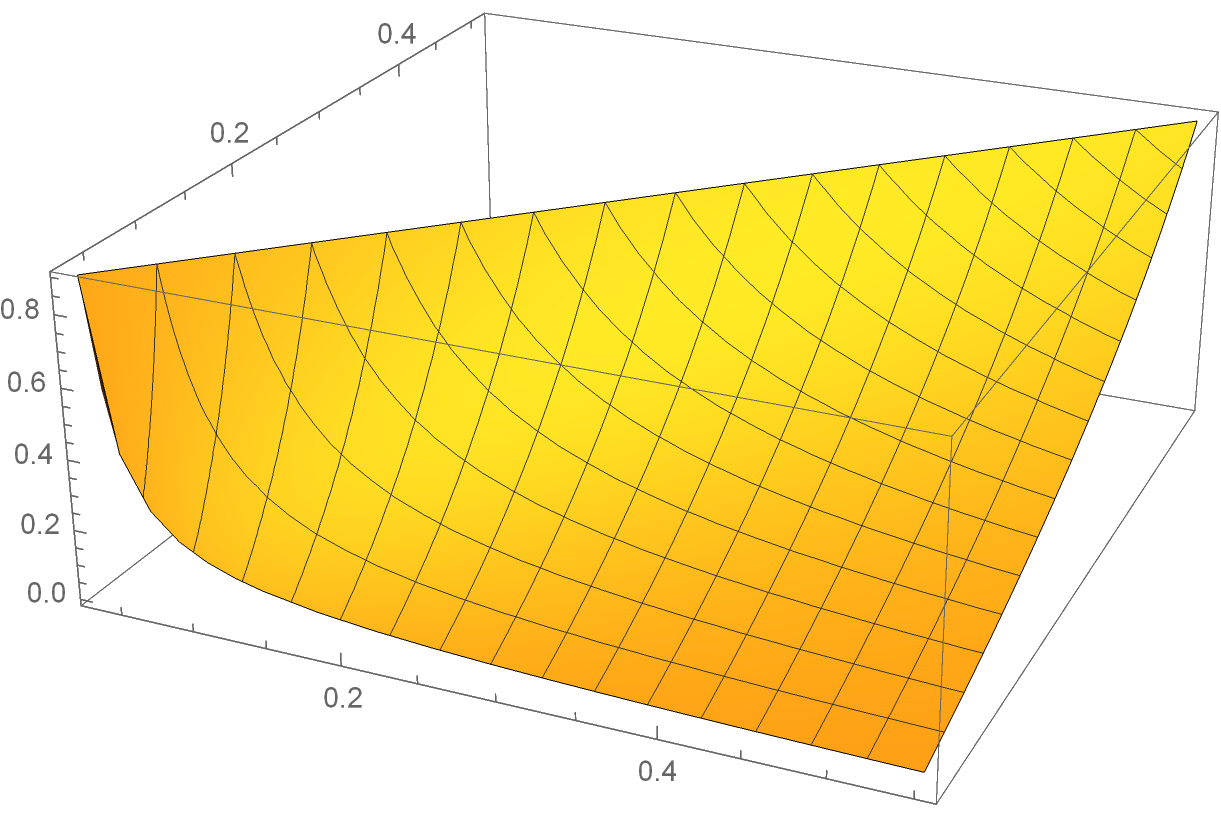}
            \put(-4,30){$\xi$}
            \put(30,3){ $x_\pi$}
            \put(10,60){ $x_B$}
        \end{overpic}

    \caption[*]{Skewness $\xi$ (\ref{skewdef}) in the $(x_\pi,x_B)$ plane at $s_{ep}=800$ GeV$^2$, $Q^2=10$ GeV$^2$, $t_\pi=-0.5$ GeV$^2$.}
    \label{skew}
\end{figure}

Let us first consider DVMP. 
We introduce the momentum fraction $x_\pi$ of the proton carried by the pion and the Bjorken variable $x_B^\pi$ for the $\gamma^*+\pi^{+*}$ subprocess 
\beq
x_\pi = \frac{p_\pi\cdot l}{p\cdot l}, \qquad x_B^\pi = \frac{Q^2}{2p_\pi \cdot q}.
\eeq
 In the limit $Q^2\to \infty$, the usual Bjorken variable $x_B$ factorizes as \cite{Amrath:2008vx}
\beq
x_B  = \frac{Q^2}{2p \cdot q} \approx x_\pi x_B^\pi .
\eeq
At low-energy, near the threshold
$s_\pi^{th}\lesssim s_\pi \ll Q^2$, an additional condition $x_B^\pi \approx 1$ follows from the relation 
\beq
s_\pi = Q^2\left(\frac{1}{x_B^\pi}-1\right)+t. 
\eeq
 Therefore, the threshold region sits along the diagonal line $x_B\approx x_\pi$ in the $(x_\pi,x_B)$ plane. There the  skewness variable is order unity
\beq
\xi
\approx \frac{x_B^\pi}{2-x_B^\pi\left(1-\frac{t_\pi}{Q^2}\right)} \approx\frac{x_B}{2x_\pi - x_B\left(1-\frac{t_\pi}{Q^2}\right)} \sim {\cal O}( 1),
\label{skewdef}
\eeq
where we ignored ${\cal O}(1/Q^2)$ terms except for the dependence on the pion momentum transfer $t_\pi=(p_\pi-p'_\pi)^2$ which can become large in actual experiments. This is plotted in Fig.~\ref{skew}. The ranges of $x_\pi,x_B$ depend on $s_{ep}= (p+l)^2$ and the upper cutoff $|t|<|t_{max}|$  \cite{Sullivan:1971kd,Amrath:2008vx}, see \cite{supple}. Working in the small-$x_\pi$ region  helps to avoid the $\Delta$-resonance region $M_{\pi n}^2=(p_n+p'_\pi)^2\sim 1.5$ GeV$^2$ \cite{Amrath:2008vx}. 

In the case of $J/\psi$  photoproduction $Q^2=0$,  we remove the electron vertex and  consider the reduced cross section $\gamma +p\to J/\psi+\pi^++n$.  Again, $\xi$  becomes order unity near the threshold $s_{\pi}\gtrsim (m_\pi+m_{J/\psi})^2\approx 10.5$ GeV$^2$. 
It is important to note that, due  to the light pion mass,  $\xi$ can be tuned closer to unity  for the pion target than for the nucleon target  in both processes.

{\it Threshold approximation}---The basic ingredients of the QCD factorization approach are the convolution integrals of GPDs. To leading order they have the  form  
\beq
{\cal H}^a(\xi,t,\mu^2) = \int_{-1}^1 dx\frac{1}{\xi-x-i\epsilon}\begin{cases} \frac{1}{2}H^{q(+)}(x,\xi,t,\mu^2) \\ \frac{1}{x}H^g(x,\xi,t,\mu^2) ,\end{cases} \label{integral}
\eeq
where $q=u,d,s,\cdots$ is the flavor label and $H^{q(+)}(x,\xi)=H^q(x,\xi)-H^q(-x,\xi)$ is the C-even part. $\mu^2$ is the renormalization scale. 
 Typically, at high energy $s_\pi\gg Q^2$ where $\xi\ll 1$, one tries to extract GPDs $H$ from ${\cal H}$ by deconvoluting the $x$-integral.  However,  it has been originally observed in \cite{Hatta:2021can,Guo:2021ibg} and further refined in \cite{Guo:2023qgu} that, when $\xi \sim {\cal O}(1)$ and for the gluon channel $a=g$, (\ref{integral})  can be well approximated by 
\begin{align} 
{\cal H}^a_{\rm trunc}(\xi,t,\mu^2)&= \frac{2}{\xi^2} \frac{5}{4} (A^{a}(t,\mu^2)+\xi^2D^a(t,\mu^2)), \label{trunc}
\end{align} 
where $A,D$ are the gravitational form factors 
\beq
A^{a}(t,\mu^2)+\xi^2 D^a(t,\mu^2)=\int_{-1}^1 dx \begin{cases}  xH^{q}(x,\xi,t,\mu^2)  \\ 
 \frac{1}{2} H^g(x,\xi,t,\mu^2). \end{cases}
\eeq
(\ref{trunc}) is nothing but the leading order term in the conformal partial wave expansion \cite{Mueller:2005ed,Muller:2013jur} of (\ref{integral}). 
Drastic as it may seem, the validity of this `threshold approximation'  for the nucleon target has been tested to next-to-leading order (NLO) in  $\phi$-electroproduction  \cite{Hatta:2025vhs} and $J/\psi$-photoproduction \cite{Guo:2025jiz}. The approximation works fine  already for moderately large values $\xi \gtrsim 0.4$. Below we test this approach for the pion target.

For this purpose, we adopt  the `algebraic' model of the pion GPDs $H^a_\pi(x,\xi,t_\pi,\mu^2)$ constructed in \cite{Chavez:2021llq}. In the Supplemental Material \cite{supple}, we discuss another model to check the model dependence. A notable feature of this model is that it contains the so-called D-term \cite{Polyakov:1999gs} which has support in the central region $|x|<\xi$ 
\beq
H_\pi^a(x,\xi,t_\pi,\mu^2)\sim \theta(\xi-|x|)D_\pi^a(x/\xi,t_\pi,\mu^2).
\eeq
Remarkably, in the chiral limit of QCD, the forward value $D_\pi^a(x/\xi,t_\pi=0)$ is strongly constrained by the soft pion theorem for all parton species $a=q,g$ \cite{Polyakov:1998ze,Chavez:2021llq}. This in particular means that  
\beq
D_\pi^q(0,\mu^2) =\int_{-1}^1  dz z D^q_\pi(z,0,\mu^2)= -A^q_\pi(0,\mu^2), 
\eeq
and similarly for the gluon. 
The model also assumes isospin symmetry which means  $H_\pi^{u(+)}(x,\xi,t_\pi)=H_\pi^{d(+)}(x,\xi,t_\pi)$ exactly.
In real QCD, there are deviations from these relations. 

\begin{figure}[t]
    \centering
 \vspace{3mm}
 \begin{overpic}[width=0.8\linewidth]{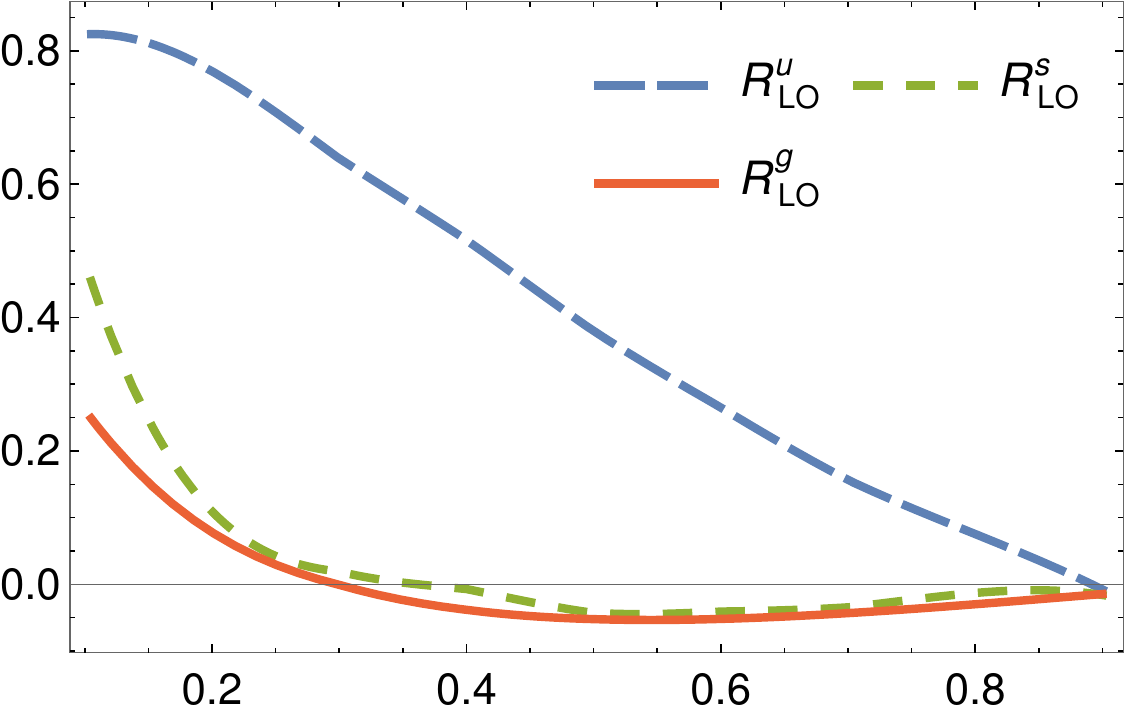}
     \put(50,-4){$\xi$}
    \end{overpic}
    \vspace{3mm}
    \caption{Truncation error $R^{a=u,s,g}$ (\ref{error}) of the leading order amplitude (\ref{integral}) at  $\mu^2=10 \, \text{GeV}^2$. }
    \label{r}
\end{figure}

Using the algebraic model evolved  to $\mu^2=10$ GeV$^2$ and setting $t_\pi=0$ for simplicity, we numerically evaluate (\ref{integral}) 
 and compare the results with their truncated versions (\ref{trunc}). 
This is plotted in Fig.~\ref{r} in terms of  the ratio 
\beq
R_f^a(\xi) = 1-\frac{\sqrt{({\rm Re}\, {\cal H}_f^a)^2+({\rm Im}\, {\cal H}_f^a)^2}}{{\cal H}^a_{f, \rm trunc}}, \label{error}
\eeq
which is the measure of the truncation error. The index $f = {\rm LO}, \, {\rm NLO}$ denotes whether the leading order or next-to-leading order hard kernel has been used. 
We immediately notice that the approximation  fails for  $u,d$ quarks, which are important for $\rho^0,\omega$-productions, unless one reaches $\xi \gtrsim 0.7$.  The main reason is that the corresponding GPD is well supported in the DGLAP region $\xi<|x|<1$ and takes a sizable value at the diagonal point $x= \xi$, giving rise to a large imaginary part ${\rm Im}\, {\cal H}_\pi^{u,d}(\xi)\propto H^{u,d}_\pi(\xi,\xi)$. This is related to the fact that  $u,d$ quarks constitute the valence quarks of the $\pi^+$. In the asymptotic limit $\mu^2\to \infty$, $H^{u,d}_\pi(\xi,\xi,\mu^2)$ should go to zero and the threshold approximation becomes exact. But this occurs at much higher scales than $\mu^2=10$ GeV$^2$ in the present model. 

On the other hand, the approximation works very well in the gluon and strangeness channels. The truncation error is less than 10\% already for $\xi > 0.2$. In contrast to $H_\pi^{u,d}$, $H_\pi^{s,g}$ are dominantly supported  in the central region $|x|<\xi$, leading to a small imaginary part in the amplitude  ${\rm Im }\, {\cal H}_\pi^{s,g}(\xi)\propto H_\pi^{s,g}(\xi,\xi)$. 
We therefore focus on  $\phi$-electroproduction and $J/\psi$-photoproduction in what follows 
and test the validity of the threshold approximation to NLO. For $\phi$-electroproduction, we use the full NLO scattering amplitude from  \cite{Muller:2013jur,Duplancic:2016bge,Cuic:2023mki}, and its truncated version  in  \cite{Hatta:2025vhs}. In the $J/\psi$ case, we use the full NLO result from \cite{Ivanov:2004vd} and its truncated version  \cite{Guo:2025jiz,supple}.  As for meson production mechanisms, we use the asymptotic distribution amplitude for $\phi$-production  and the leading order non-relativistic QCD (NRQCD) vertex for $J/\psi$ production (see \cite{Hatta:2025vhs,Guo:2025jiz}). These approximations can be systematically improved in future.

\begin{figure}
    \centering
    \begin{overpic}[width=0.8\linewidth]{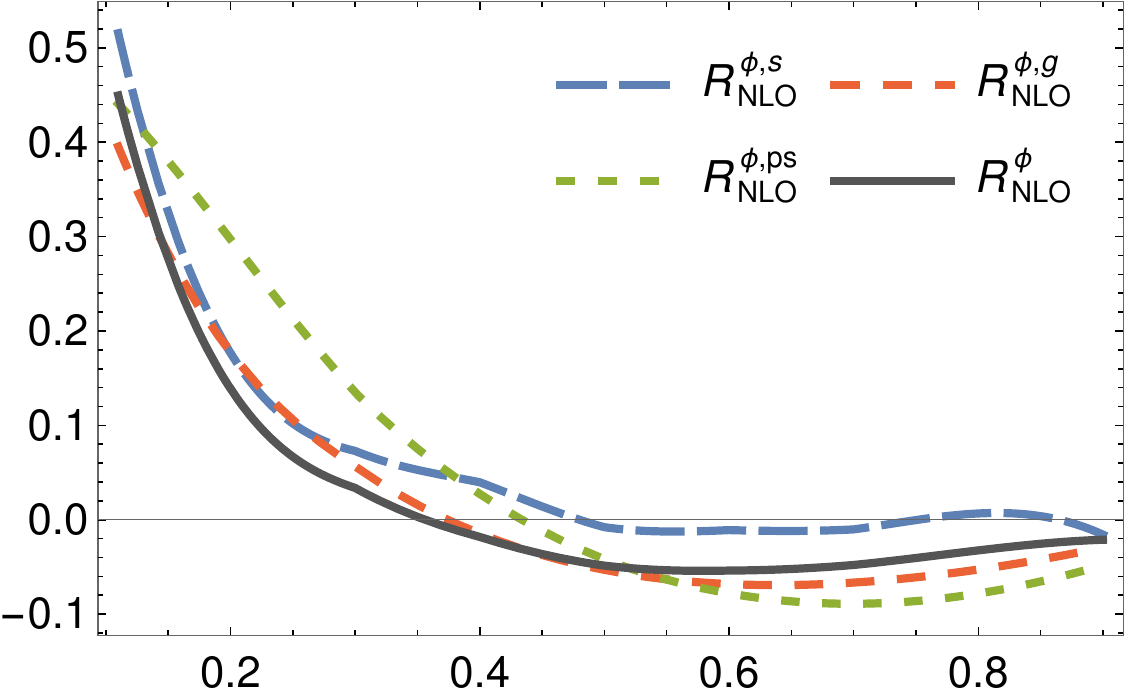}
     \put(50,-4){$\xi$}
    \end{overpic}
    \vspace{6mm}
    \\
 \begin{overpic}[width=0.8\linewidth]{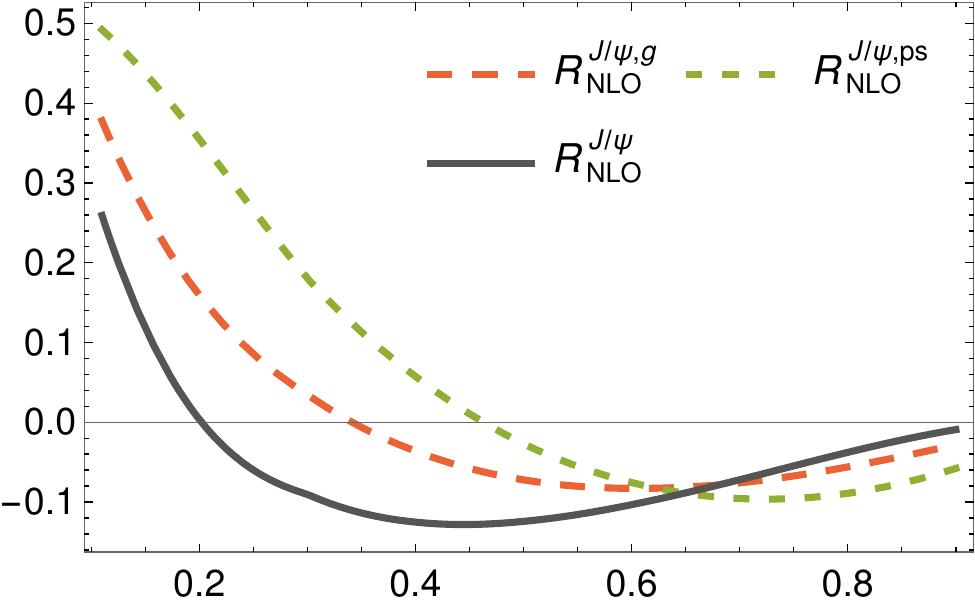} 
  \put(50,-4){$\xi$}
     \end{overpic}
     \vspace{2mm}
\caption{Truncation error (\ref{error}) in the full NLO amplitude at $\mu^2=10$ GeV$^2$. Top: $\phi$-electroproduction. Bottom: $J/\psi$-photoproduction. }
\label{phi}
\end{figure}

The results for $R_{\rm NLO}$ are plotted in Fig.~\ref{phi}. Each plot shows $R_{\rm NLO}$ for different partonic channels $a=g,s,ps$ where  $ps$ means the `pure singlet' contribution (the sum of GPDs/GFFs over all flavors) that appears at NLO. The total NLO amplitude is indicated by the solid line.  The threshold approximation is excellent $|R_{\rm NLO}^\phi|<0.05$ for $\phi$-electroproduction at $\mu^2=10\, {\rm  GeV}^2=Q^2$ (upper plot). As for  $J/\psi$-photoproduction, the approximation is not as good as in the $\phi$-production case due to the partial cancellation between the quark and gluon terms. Still, it is decent $|R_{\rm NLO}^{J/\psi}|<0.15$  and increasingly gets better as $\xi\to 1$.

{\it Predictions at the EIC and JLab}---Now that we have demonstrated the validity of the threshold approximation  for $\phi$-electroproduction and $J/\psi$-photoproduction, we make concrete NLO predictions for the JLab and EIC kinematics. The relevant cross section formulas are summarized in  \cite{supple}. 
We assume the monopole and dipole form for the $A,D$ form factors \cite{Tanaka:2018wea} (see also \cite{Tong:2021ctu})
\beq
A_\pi^{a}(t,\mu^2)= \frac{A_\pi^{a}(0,\mu^2)}{1-t/m_A^2}, \quad D_\pi^{a}(t,\mu^2)= \frac{D_\pi^{a}(0,\mu^2)}{(1-t/m_D^2)^2}. \label{mono}
\eeq
We use $m_A=1.6$ GeV and $m_D=1.1$ GeV as in \cite{Hatta:2025vhs}, but ultimately these parameters should be determined from future experimental data. 
The forward values $A_\pi^a(0,\mu^2)$  are constrained by the pion PDF. We use $A_\pi^u=A_\pi^d= 0.28$ (isospin symmetry), $A_\pi^s=0.02$ and  $A_\pi^g=0.41$ at the reference scale $\mu^2=2$ GeV$^2$ from the algebraic model and  consider their one-loop evolution with $n_f=4$ flavors. We impose the constraint $D_\pi^a(0,\mu^2)=-A_\pi^a(0,\mu^2)$ from the soft pion theorem, although possible deviations from this formula are interesting and can be  implemented in a straightforward manner.

\begin{figure}
\centering
        \begin{overpic}[width=0.8\linewidth]{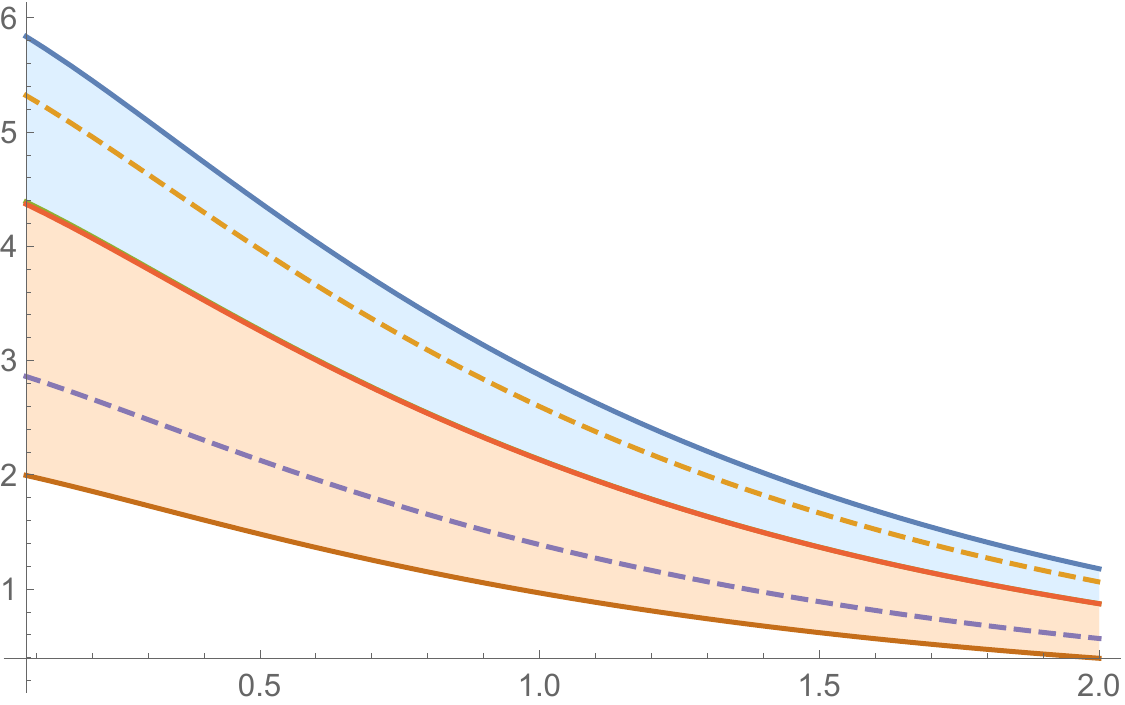}
         \put(-8,10){\rotatebox{90}{ $d\sigma^{ep}_L/dt_\pi\;(\text{pb}/\text{GeV}^4)$}}
            \put(40,-6){ $|t_\pi|$ GeV$^2$}
        \end{overpic} \\
        \vspace{6mm}
\begin{overpic}
[width=0.8\linewidth]{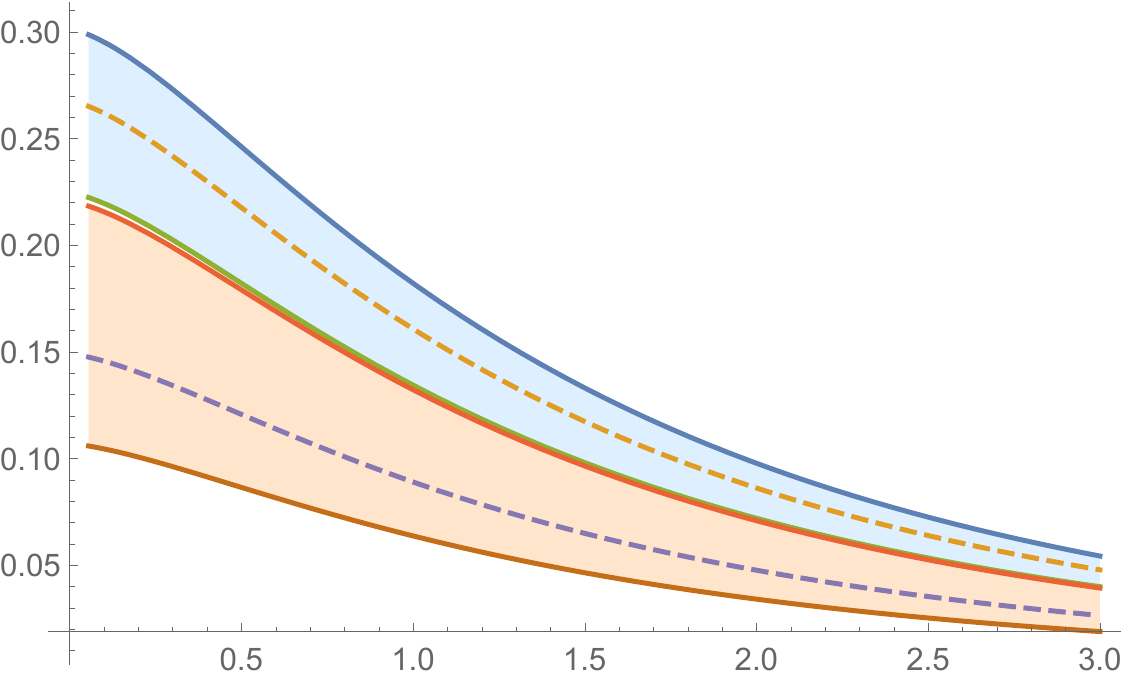}
         \put(-8,10){\rotatebox{90}{$d\sigma^{ep}_L/dt_\pi\;(\text{pb}/\text{GeV}^4)$}}
            \put(40,-6){$|t_\pi|$ GeV$^2$}
\end{overpic}
\vspace{5mm}
    \caption[*]{Longitudinal part of the $ep$ cross section (\ref{epcross}) at $x_\pi=0.3$, $x_B=0.2$ as a function of $|t_\pi|$ at the EIC. Top: $Q^2=5$ GeV$^2$, Bottom: $Q^2=10$ GeV$^2$.  Blue and orange band are the NLO and LO cross sections, respectively, varied within $Q/2<\mu < 2Q$. }
    \label{dtpi}
\end{figure}

First consider $\phi$-electroproduction  at the EIC with  $s_{ep}=800$ GeV$^2$.  
We write the $e+p$ cross section as (not to be confused with the reduced $\gamma^*+p$ cross section $d\sigma_{T/L}^{\gamma^* p}$) \newpage
\begin{align}
\frac{d\sigma^{ep}}{dx_B dQ^2 dt_\pi  dx_\pi }\equiv 
\frac{d\sigma^{ep}_T}{dx_B dQ^2 dt_\pi  dx_\pi } +\varepsilon\frac{d\sigma^{ep}_L}{dx_B dQ^2 dt_\pi  dx_\pi }, \label{epcross}
\end{align}
where the longitudinal/transverse photon flux ratio $\varepsilon$ is  
\beq
\varepsilon = \frac{1-y_\pi}{1-y_\pi + \frac{y_\pi^2}{2}} ,\qquad  y_\pi= \frac{q\cdot p_\pi}{l\cdot p_\pi}.
\eeq
 In the present kinematics, $y_\pi$ is approximately equal to  the usual DIS variable $y$  \cite{Amrath:2008vx}
 \beq
 y_\pi \approx y=\frac{q\cdot p}{l\cdot p}= \frac{Q^2}{(s_{ep}-m_N^2)x_B}, \label{rose}
 \eeq
  where $m_N$ is the nucleon mass.
We focus on the longitudinal part $d\sigma_L^{ep}$ which admits QCD factorization \cite{Collins:1996fb}. 
Fig.~\ref{dtpi} shows $d\sigma_L^{ep}$ evaluated at a  representative point $(x_\pi,x_B)=(0.3,0.2)$ with two different values $Q^2=5$ GeV$^2$ (top) and $Q^2=10$ GeV$^2$ (bottom). The NLO cross section (blue band) is larger and has smaller scale uncertainties than the LO cross section (orange band). In these plots,  $\xi> 0.5$, where  the truncation error (see Fig.~\ref{phi}) is expected to be smaller than the scale uncertainties and possible higher twist effects of order $t_\pi/Q^2$. The cross section is dominated by the gluon exchange which is partly canceled by the NLO quark contribution. Because of the soft pion theorem $D_\pi^s(0)\approx -A_\pi^s(0)$, the large-$D_s$ scenario for the nucleon considered in \cite{Hatta:2021can,Hatta:2025vhs} is unlikely for the pion.  
The rather weak $t_\pi$-dependence is due to the monopole ansatz (\ref{mono}) which can be experimentally tested. 

The corresponding measurement at JLab will be more difficult, as the allowed ranges of $x_\pi,x_B$ \cite{supple} 
are narrower at the current JLab energy  $s_{ep}\approx 23$ GeV$^2$. One is forced to go to lower regions in $Q^2$ at the risk of reducing the accuracy of perturbative predictions. The proposed 22 GeV upgrade ($s_{ep}\approx 42$ GeV$^2$)  \cite{Accardi:2023chb}, if realized, certainly improves the situation.  The Electron-Ion Collider in China (EIcC) \cite{Anderle:2021wcy} offers another possibility. One advantage of low energy experiments 
is that it may be possible to perform the longitudinal/transverse (L/T) decomposition via the `Rosenbluth separation' method. Usually, one varies   $\varepsilon$ by changing the beam energy $s_{ep}$ at fixed $Q^2,x_B$ (see (\ref{rose})). But in the Sullivan process, one may as well fix $s_{ep}$ and vary $x_B$ along the `degenerate' line $x_B^\pi \approx x_B/x_\pi=const$. This allows one to  access the low-$\varepsilon$ region critical for the method \cite{JeffersonLab:2008gyl}.

\begin{figure}[t]
\centering
        \begin{overpic}[width=0.4\textwidth]{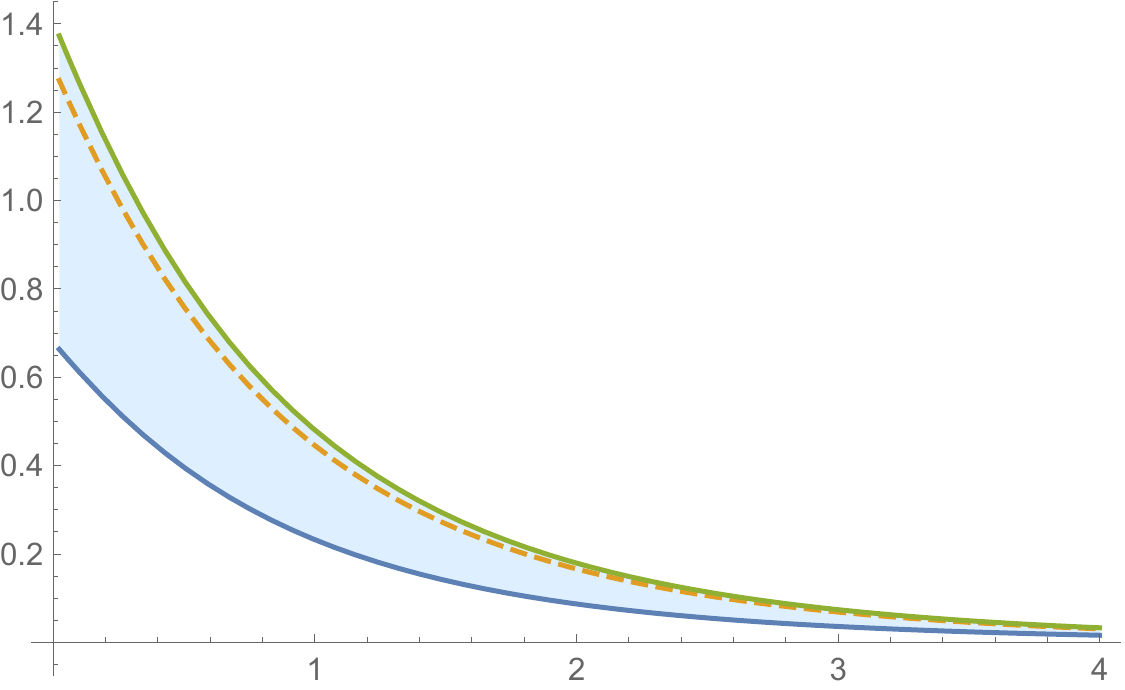}
            \put(-9,9){\rotatebox{90}{$d\sigma^{\gamma p}/dx_\pi dt_\pi$ (nb/GeV$^2$) }}
            \put(50,-5){ $|t_\pi|$ GeV$^2$}
        \end{overpic}\vspace{6mm}
        \begin{overpic}[width=0.4\textwidth]{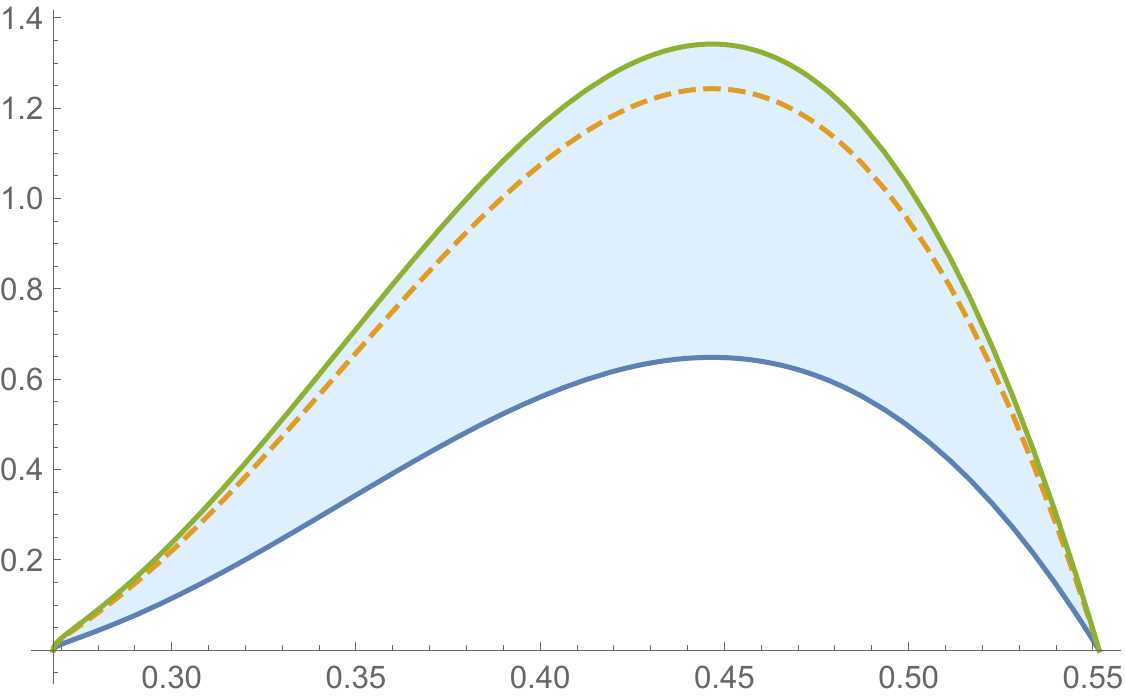}
            \put(-8,20){\rotatebox{90} {$d\sigma^{\gamma p}/dx_\pi$ (nb)} }
            \put(53,-4){ $x_\pi$}
        \end{overpic}
        \vspace{2mm}
  \caption[*]{Upper plot:  $J/\psi$-photoproduction cross section (\ref{jpsi2D}) at JLab with  $s_{\gamma p}=40$ GeV$^2$ and $x_\pi=0.45$. Lower plot:  Cross section integrated over $t_\pi$. The three curves correspond to  $\mu=2m_{J/\psi},m_{J/\psi}$ and $m_{J/\psi}/2$ from top to bottom. }
     \label{22}
\end{figure}

Next we consider $J/\psi$-photoproduction at JLab.
We compute the reduced $\gamma+p$ cross section 
\beq
\frac{d\sigma^{\gamma p}}{dx_\pi dt_\pi}, \label{jpsi2D}
\eeq
treating the photon energy $q^0=E_\gamma$  as a  fixed parameter. This can be controllably  done as in the recent JLab measurements of $J/\psi$ off the nucleon target \cite{GlueX:2019mkq,Duran:2022xag,GlueX:2023pev}. However, the same  is challenging for the pion target in the Sullivan process due to the lack of  phase space. We thus assume the 22 GeV upgrade \cite{Accardi:2023chb} and fix $s_{\gamma p}  = (q + p)^2 =40$ GeV$^2$.  
In Fig.~\ref{22}, we plot (\ref{jpsi2D}) 
at $x_\pi=0.45$ corresponding to $\xi>0.38$.  The uncertainty band corresponds to the window $m_{J/\psi}/2<\mu < 2m_{J/\psi}$. We observe that the region $\mu\sim m_{J/\psi}/2\approx m_c$ suffers from a rather strong scale dependence which we attribute to the cancellation between the LO gluon and NLO quark contributions. This can be avoided by selecting $\mu> m_{J/\psi}$  where the NLO cross section shows a fairly stable plateau in $\mu$ \cite{supple}.   The magnitude of the cross section is of the same order as in the nucleon target production \cite{GlueX:2019mkq,Duran:2022xag,GlueX:2023pev}  and falls slower with $|t_\pi|$. The lower plot is the $t_\pi$-integrated cross section which clearly shows the rise of the cross section above the threshold $x_\pi>0.27$. The upper limit $x_\pi\lesssim 0.55$ is an artificial cutoff imposed by the condition $|t|<|t_{max}|=0.6$ GeV$^2$  \cite{Amrath:2008vx,Qin:2017lcd}.

{\it Conclusions}---In this paper we have proposed a novel method to experimentally constrain the gluon and strangeness  gravitational form factors of the pion. We have demonstrated to NLO in QCD that $\phi$-electroproduction and $J/\psi$-photoproduction near the threshold have strong sensitivities to the pion GFFs. Moreover, thanks to the light pion mass, one can explore larger $\xi$ and lower $|t_\pi|$ regions than for the nucleon target, which makes theoretical treatments cleaner. The predicted NLO cross sections are well in the measurable range with the high luminosities of the EIC and JLab. 

There are many directions for future  exploration. First, it would be  interesting to further investigate the sensitivity of near-threshold $\rho^0$ and $\omega$-productions to the up and down quark GFFs.  The region  $\xi >0.7$ (see Fig.~\ref{r}) is not inaccessible for the pion target, but there are practical issues such as model dependence and experimental feasibility. Second, while near-threshold photoproduction is challenging to measure at the EIC, electroproduction of $J/\psi$ should be possible \cite{Boussarie:2020vmu}. The calculation can be pushed to NLO using the result of  \cite{Flett:2021ghh}.  
One can also consider photoproduction in proton-nucleus ultraperipheral collisions (UPCs)  \cite{Hatta:2019lxo}. 
The Sullivan process in these reactions has been studied in the high energy limit $s_\pi \to \infty$ \cite{Goncalves:2015mbf,Goncalves:2016uhj} but not in the threshold region. 
Finally, the need for reconstructing the scattered particles $e,V,\pi^+,n$ poses significant experimental challenges. Careful investigations based on realistic experimental setups and event generator simulations (see, e.g., \cite{Hatta:2025vhs}) are crucial in this respect. We hope our work triggers future efforts  in this direction. \\

{\it Acknowledgments}---We thank Henry Klest and Kornelija Passek-K. for the collaboration in \cite{Hatta:2025vhs} and  discussions on various  aspects of this work.   We also thank Jose Morgado-Chavez and Cedric Mezrag for numerical tables of the GPD models in \cite{Chavez:2021llq}. 
We were supported by the U.S. Department
of Energy under Contract No. DE-SC0012704, and also by LDRD funds from Brookhaven Science Associates.

\bibliography{ref}

\pagebreak
\widetext
\begin{center}
\textbf{\large Supplemental Materials}
\end{center}
\setcounter{section}{0}
\setcounter{equation}{0}
\setcounter{figure}{0}
\setcounter{table}{0}
\setcounter{page}{1}
\makeatletter
\renewcommand{\theequation}{S\arabic{equation}}
\renewcommand{\thefigure}{S\arabic{figure}}

%

\onecolumngrid

\section{Kinematics and cross section formulas}

\subsection{Meson electroproduction}
Here we review the kinematics of DVMP in the Sullivan process  building on  \cite{Sullivan:1971kd,Amrath:2008vx}.
We introduce the usual DIS variables  
\beq
s_{ep}=(p+l)^2, \qquad x_B=\frac{Q^2}{2p\cdot q}, \qquad y=\frac{q\cdot p}{l\cdot p}
\eeq
and their counterparts in the $e+\pi^{+*}$ subprocess 
\beq
 x_B^\pi =  \frac{Q^2}{2p_\pi\cdot q},\qquad y_\pi = \frac{q\cdot p_\pi}{l\cdot p_\pi}.
\eeq
 where $p_\pi = p-n$.  We treat the electron as massless. 
The other important variables  are the momentum fraction $x_\pi$ of the proton carried by the pion  and the momentum transfer $t_\pi$ in the subsequent scattering $\gamma^*(q)+\pi^{+*}(p_\pi)\to V+\pi^+(p'_\pi)$
\beq
x_\pi = \frac{p_\pi \cdot l}{p\cdot l}, \qquad  
t_\pi= (p_\pi - p'_\pi)^2.
\eeq
In the $Q^2\to \infty$ limit, one generically has \cite{Amrath:2008vx}
\beq
x_B^\pi \approx \frac{x_B}{x_\pi}, \qquad y_\pi \approx y =\frac{Q^2}{(s_{ep}-m_N^2)x_B}. 
\eeq

The nucleon momentum transfer $t=p_\pi^2$ is restricted to
\beq
|t_{min}|= \frac{x_\pi^2m_N^2}{1-x_\pi} < |t| < |t_{max}| ,
\eeq
The lower limit is a kinematical constraint. The upper limit $t_{max}$ ensures that the transition $p\to n$ is dominated by the one-pion emission. Typically $|t|\ll 1$ GeV$^2$, but we allow for   $|t_{max}|=0.6$ GeV$^2$ \cite{Amrath:2008vx,Qin:2017lcd}.
The range of  $x_\pi$ is
\beq
 \frac{s_\pi^{th} +Q^2}{(s_{ep}-m_N^2)y_{max}} \lesssim x_\pi  < \frac{-|t_{max}|+\sqrt{t_{max}^2+4|t_{max}|m_N^2}}{2m_N^2} \approx 0.55, \label{xpimax}
\eeq
where $y_{max}$ is the maximal value of $y$ in a given experiment.    The upper limit  follows from the condition $|t_{min}|<|t_{max}|$ and the lower limit is from 
\beq
 x_\pi y \approx x_\pi y_\pi = \frac{2p_\pi\cdot q}{2p\cdot l}  =\frac{s_\pi +Q^2-t}{s_{ep}-m_N^2}.
\eeq
In principle, $y_{max}\approx 1$, but in practice a smaller value is adopted to suppress QED radiative corrections which become important when $y\to 1$. We use $y_{max}=0.85$. 
Finally, the constraint on $x_B$ can be deduced from the formula \cite{Amrath:2008vx} 
\beq
s_\pi 
&= & Q^2\left(\frac{x_\pi}{x_B}-1\right)+2yx_Bx_\pi m_N^2+(1-yx_B)t \nn 
&&-2\cos(\phi_{l'}-\phi_{n})\sqrt{(1-y)Q^2-(yx_Bm_N)^2}\sqrt{(1-x_\pi)(t_{min}-t)} \label{spi}
\eeq
To leading order in $Q^2$, the cross section does not depend on the azimuthal angles $\phi_{l'},\phi_n,\phi_\pi$. The cosine term can then be averaged out. Besides, from (\ref{xpimax}), 
\beq
x_Bx_\pi m_N^2 <  x_\pi^2 m_N^2 < |t_{max}|,
\eeq
is small compared to $s_{\pi}>s^{th}_\pi\sim 1$ GeV$^2$.
We thus find 
\beq
\frac{Q^2}{(s_{ep}-m_N^2) y_{max}} < x_B \lesssim {\rm min}\left[\frac{x_\pi}{1+\frac{s_\pi^{th}}{Q^2}},\frac{x_\pi}{1-\frac{t_\pi}{2Q^2}} \right]. \label{xbmax}
\eeq
The second upper limit comes from the condition $\xi<1$  which is relevant only at high momentum transfer $|t_\pi|>2s_\pi^{th}$.

 The fully-differential $e+p\to e'+V+\pi^++n$ cross section takes the form \cite{Amrath:2008vx}
\beq
\frac{d\sigma^{ep}}{dydQ^2 d\phi_{l'}dt_\pi d\phi_\pi dt dx_\pi d\phi_n}= \frac{(\sqrt{2}g_{\pi NN}F(t))^2}{128(2\pi)^8(s_{ep}-m_N^2)\sqrt{(s_\pi+Q^2+t)^2-4s_\pi t}} \frac{-t}{(m_\pi^2-t)^2}|{\cal M}_V(\xi,t_\pi)|^2 \label{cross}
\eeq
where $g_{\pi NN}\approx 13.1$ is the pion-nucleon coupling and $F(t)$ with $F(m_\pi^2)=1$ is the associated form factor. ($\sqrt{2}$ is the  isospin factor.)  We decompose the hard cross section into the longitudinal and transverse parts 
\beq
|{\cal M}_V(\xi,t_\pi)|^2 =  \frac{2e^4}{Q^2}\frac{(4\pi)^2}{1-\varepsilon}  \left(|{\cal H}_V^T(\xi,t_\pi)|^2+\varepsilon |{\cal H}_V^L(\xi,t_\pi)|^2\right)+\cdots,
\eeq
where we ignore the angular dependent terms and 
\beq
\varepsilon = \frac{1-y_\pi}{1-y_\pi + \frac{y_\pi^2}{2}} \approx \frac{1-y}{1-y + \frac{y^2}{2}},
\eeq
is the longitudinal/transverse virtual photon flux ratio. 
In DVMP, only the longitudinal part is reliably calculable. In the case of $V=\phi$, and in the threshold approximation to NLO  \cite{Hatta:2025vhs}, 
\begin{align}
{\cal H}_\phi^L(\xi,t_\pi)\approx e_s\frac{C_Ff_\phi}{N_cQ} \frac{2\alpha_s(\mu)}{\xi^2} \frac{15}{2}\Biggl[\left\{1+\frac{\alpha_s}{2\pi}\left(25.7-2n_f +\left(-\frac{131}{18}+\frac{n_f}{3}\right)\ln \frac{Q^2}{\mu^2}\right)\right\}(A_\pi^s(t,\mu)+\xi^2D_\pi^s(t,\mu)) \label{hxi} \\  
  +\frac{\alpha_s}{2\pi}\left(-2.34+\frac{2}{3}\ln \frac{Q^2}{\mu^2}\right) \sum_q (A_\pi^{q}+\xi^2 D_\pi^{q})  + \frac{3}{8}\left\{1+\frac{\alpha_s}{2\pi}\left(13.9-\frac{83}{18}\ln \frac{Q^2}{\mu^2}\right)\right\} (A_\pi^g+\xi^2D_\pi^g) \Biggr],\nonumber 
\end{align}
where $e_s=-\frac{1}{3}$, and 
$f_\phi$ is the  decay constant. We have assumed the asymptotic form $\varphi_\phi(u)\approx 6u(1-u)$ for the meson distribution amplitude.  

Returning to (\ref{cross}), in the present kinematics we may approximate  
\beq
\frac{1}{\sqrt{(s_\pi+Q^2+t)^2-4s_\pi t}}\approx \frac{x_B}{x_\pi Q^2}.
\eeq
We can then trivially perform all the angular integrals. As for the $t$-integral, we follow   \cite{Amrath:2008vx} and introduce  
\beq
\Pi(x_\pi,t_{max}) = x_\pi \frac{g_{\pi NN}^2}{8\pi^2}\int^{t_{min}}_{t_{max}} dt F^2(t)\frac{-t}{(m_\pi^2-t)^2}. \label{pi}
\eeq 
We use the same  monopole model $F(t)= \frac{\Lambda^2-m_\pi^2}{\Lambda^2-t}$ with $\Lambda=0.8$ GeV as in \cite{Amrath:2008vx}. 
Then the $t$-integral can be done explicitly. 
 Switching variables $y\to x_B$,  
we arrive at 
\beq
\frac{d\sigma^{ep}}{dx_B dQ^2 dt_\pi  dx_\pi } 
&=&\frac{2\pi \alpha_{em}^2\Pi(x_\pi,t_{max})}{x_\pi^2x_B(s_{ep}-m_N^2)^2Q^2(1-\varepsilon)}  \left(|{\cal H}_V^T(\xi,t_\pi)|^2+\varepsilon |{\cal H}_V^L(\xi,t_\pi)|^2\right) \nn 
&\equiv& \frac{d\sigma^{ep}_T}{dx_B dQ^2 dt_\pi  dx_\pi } +\varepsilon\frac{d\sigma^{ep}_L}{dx_B dQ^2 dt_\pi  dx_\pi } . \label{total}
\eeq
Note that $d\sigma_L^{ep}$  contains the usual factor $\frac{1}{1-\varepsilon}$ in our definition. We emphasize that  (\ref{total}) is the $e+p$ cross section directly measured in experiments. The reader should not confuse this with the common notation $d\sigma_{T/L}$  often used for the reduced $\gamma^*+p$ cross section. 

\subsection{Quarkonium photoproduction}

\begin{figure}[t]
        \begin{overpic}[width=0.46\textwidth]{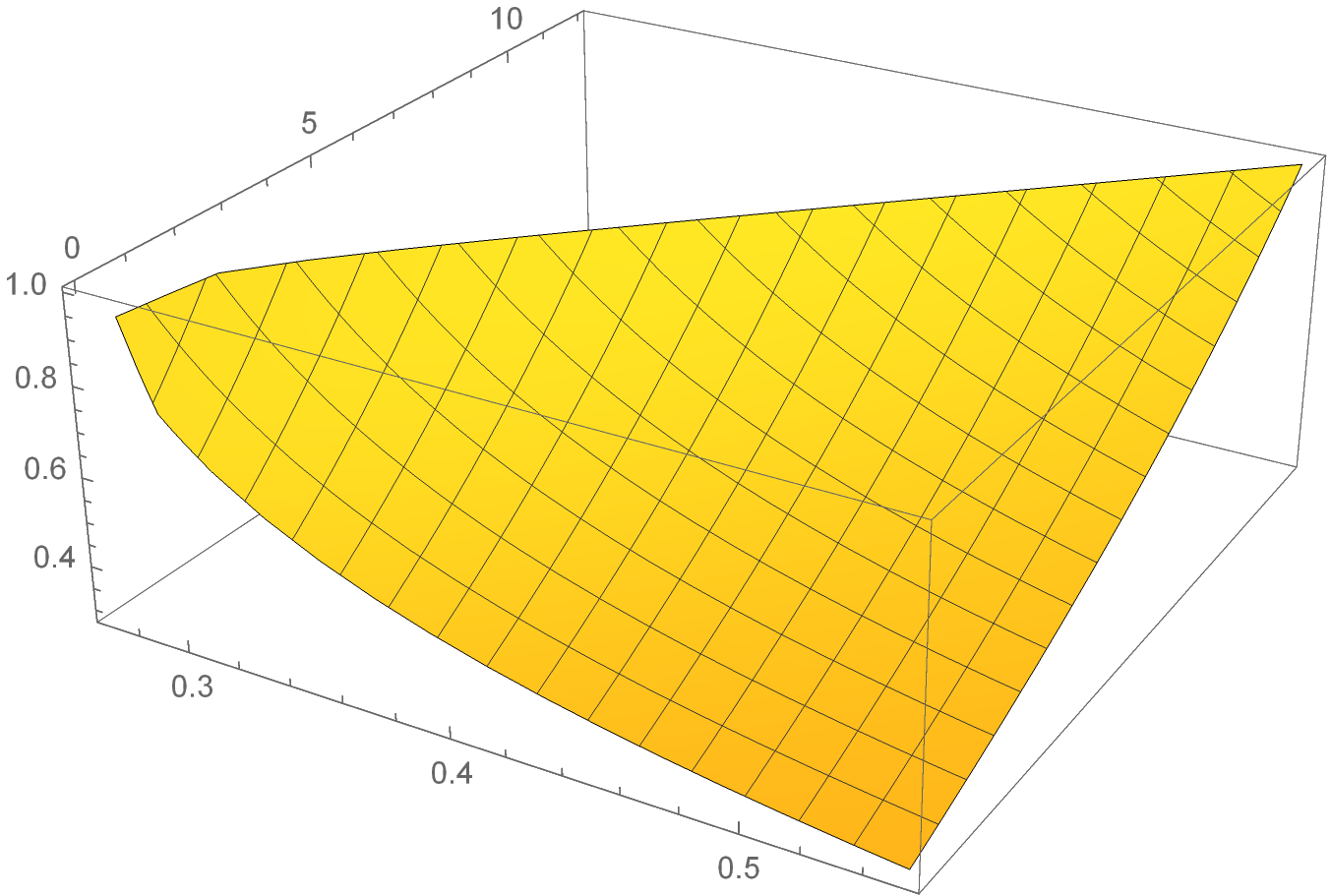}
            \put(10,57){\rotatebox{26}{$|t_\pi|$   GeV$^2$}}
            \put(30,+3){ $x_\pi$}
            \put(-5,35){$\xi$}
        \end{overpic}

    \caption[*]{ Skewness in $J/\psi$-photoproduction at $s_{\gamma p}=40$ GeV$^2$.  }
    \label{skewJ}
\end{figure}

In the case of $J/\psi$-photoproduction $Q^2=0$ at fixed $\gamma+p$ center-of-mass energy $s_{\gamma p}=(p+q)^2$, the variables $x_B,x_B^\pi$ are not needed and $x_\pi = \frac{p_\pi \cdot q}{p\cdot q}$. We have the constraint 
\beq
s_\pi = x_\pi(s_{\gamma p}-m_N^2)+t 
>(m_\pi+m_J)^2 \approx 10.5\ {\rm GeV}^2 
\eeq
which leads to 
\beq
 \frac{10.5\, {\rm GeV}^2}{s_{\gamma p}-m_N^2} \lesssim x_\pi < 0.55, \label{xpimaxJ}
 \eeq
 where the upper limit is the same as before  (\ref{xpimax}). We define skewness as ($v^{\pm} \equiv \frac{1}{\sqrt{2}}(v^0 \pm v^3)$)
\beq
\xi = \frac{p^+_\pi -p'^+_\pi}{p^+_\pi +p'^+_\pi} ,\label{skewdef2}
\eeq
in the $\gamma^*+\pi^{+*}$ center-of-mass frame. 
 Fig.~\ref{skewJ} shows $\xi$ in the $(x_\pi,|t_\pi|)$ plane with $s_{\gamma p}=40$ GeV$^2$  and $t\approx 0$. The threshold  corresponds to the lower limit $x_\pi\approx  0.27$.
 
\begin{figure}
\vspace{10mm}
     \begin{overpic}[
         width=0.5\textwidth]{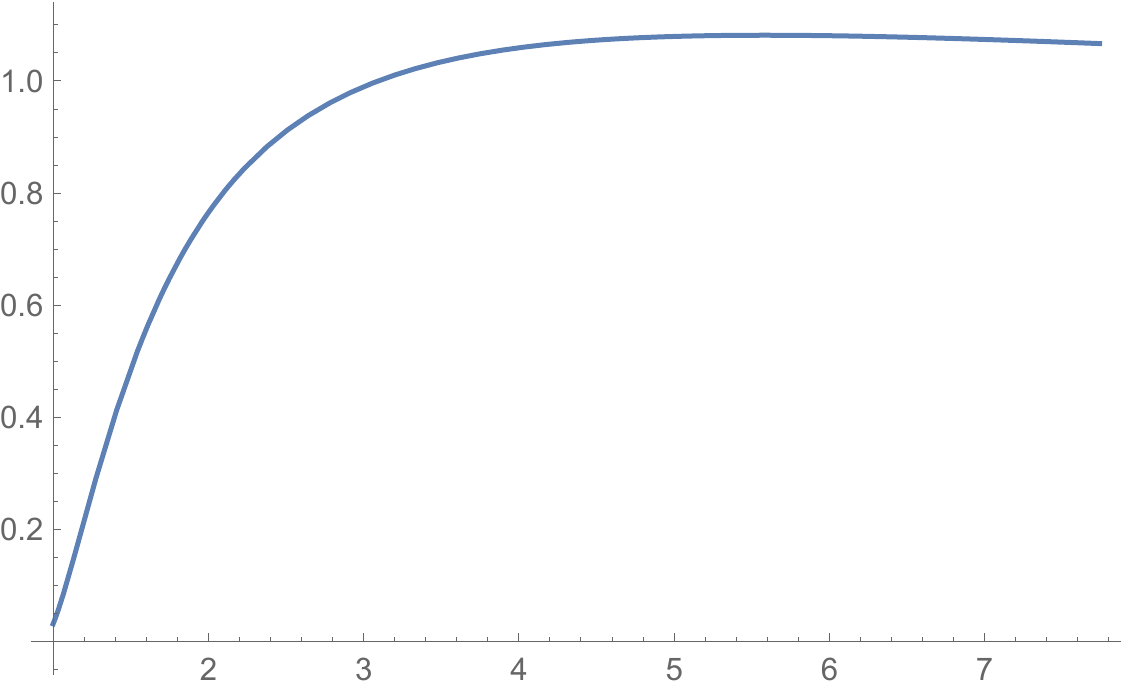}
          \put(50,-5){$\mu$ GeV}
           \put(-22,52){\Large{$\frac{d\sigma(\mu)}{d\sigma(m_{J/\psi})}$}}
     \end{overpic}
    \vspace{4mm}
    \caption{The renormalization group scale $\mu$ dependence of the  NLO $J/\psi$ cross section normalized to unity at $\mu=m_{J/\psi}$. }
    \label{ratio}
\end{figure}

Within the non-relativistic QCD (NRQCD) framework, 
the differential cross section of the reduced process $\gamma + p \to V+\pi^+ + n$ is 
\beq
&&\frac{d\sigma^{\gamma p}}{dt_\pi d\phi_\pi dt dx_\pi d\phi_n}=\frac{(\sqrt{2}g_{\pi NN}F(t))^2}{32(2\pi)^5(s_{\gamma p}-m_N^2)(s_\pi -t)}  \frac{-t}{(m_\pi^2-t)^2} 4\pi \alpha_{em}e_q^2\frac{(16\pi)^2}{N_c m_V^3}|\psi_V(0)|^2 |{\cal H}_V|^2,  
\eeq
where $|\psi_V(\vec{r}=0)|^2$ is the quarkonium wavefunction evaluated at the origin. For $J/\psi$, 
$e_q=\frac{2}{3}$ and  $|\psi(0)|^2=1.0952/(4\pi)$ GeV$^3$ \cite{Eichten:2019hbb}. 
Integrating over $t,\phi_\pi,\phi_n$, we find 
\beq
\frac{d\sigma^{\gamma p}}{dt_\pi dx_\pi } = 64\pi^2\alpha_{em}e_q^2\frac{\Pi(x_\pi,t_{max})}{ x_\pi^2(s_{\gamma p}-m_N^2)^2}\frac{ |\psi_V(0)|^2}{N_cm_V^3} |{\cal H}_V|^2. \label{quarkonium}
\eeq
To leading order, the scattering amplitude is given by the gluon GPD of the pion
\beq
{\cal H}_V(\xi,t_\pi)&=& \alpha_s\int_{-1}^1 \frac{dx}{2x} \left(\frac{1}{\xi-x-i\epsilon}-\frac{1}{\xi+x-i\epsilon}\right)H^g_\pi(x,\xi,t_\pi) + {\cal O}(\alpha_s^2) . \label{quark}
\eeq
We use the NLO ${\cal O}(\alpha_s^2)$ result from  \cite{Ivanov:2004vd}. In the threshold approximation, this reduces to 
 (see also \cite{Guo:2025jiz}) 
\begin{align}
{\cal H}_{\rm trunc}(\xi)  =\frac{2\alpha_s(\mu)}{\xi^2} \frac{5}{4} \Biggl[\left(1+\frac{\alpha_s}{2\pi} \left(5.77-\frac{11}{2}\ln \frac{4m_c^2}{\mu^2}\right)\right) (A_\pi^g+\xi^2 D_\pi^g)  + \! \frac{\alpha_s}{2\pi} \!\left(-6.94+\frac{16}{9} \ln \frac{4m_c^2}{\mu^2}\right)\sum_q(A_\pi^q+\xi^2 D_\pi^q)\Biggr], \label{nloJ}
\end{align}
where $m_c\approx m_{J/\psi}/2$ is the charm quark mass. Importantly, the NLO quark contribution is negative. 
Note that (\ref{nloJ}) is renormalization group invariant up to terms of order ${\cal O}(\alpha_s^3)$.   In Fig.~\ref{ratio}, we plot the cross section at $x_\pi=0.45$, $t_\pi=t_\pi^{min}$ as a function of $\mu$ normalized by its value at $\mu=m_{J/\psi}$. 
We see a sizable plateau in the region  $\mu\gtrsim m_{J/\psi}$ demonstrating the stability of the NLO cross section. On the other hand, 
we observe a strong $\mu$-dependence  around $\mu\sim m_c=m_{J/\psi}/2$ which we attribute to a  cancellation between the NLO quark and LO gluon terms. Thus the preferred scale choice is $\mu^2\sim (m_{J/\psi})^2\approx 10 \, {\rm GeV}^2$ or even higher, rather than $\mu^2\sim m_c^2$.

\section{Phenomenological model }

Here we consider another pion GPD model constructed in \cite{Chavez:2021llq} called the `phenomenological' model. The two models (algebraic and phenomenological) treat gluons and sea quarks at small-$x$ very differently. In the forward limit, at low renormalization scales $\mu^2$, the algebraic model features a strong rise of the gluon PDF at small-$x$ $G(x)\sim x^{-3/2}$, while in the phenomenological model the growth is minimal $G(x)\sim 1/x$. Therefore, one might expect that the actual pion GPDs fall between these  models. However, the phenomenological model is disfavored as we explain below.

\begin{figure}
    \centering
     \begin{minipage}{0.45\textwidth}
     \begin{overpic}[
         width=\textwidth]{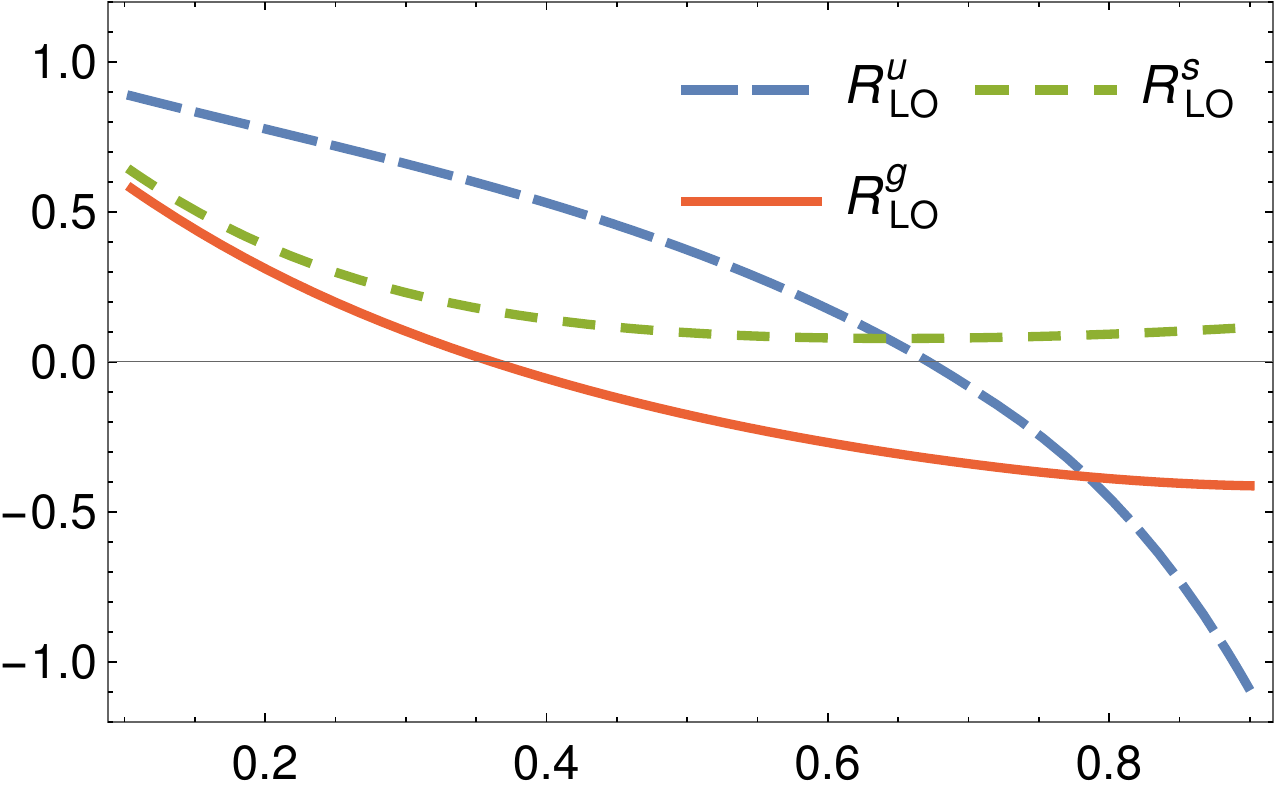}
          \put(55,-3){$\xi$}
     \end{overpic}
      \end{minipage}
      \hspace{0.05\textwidth}
    \begin{minipage}{0.45\textwidth}
    \begin{overpic}[
         width=\textwidth]{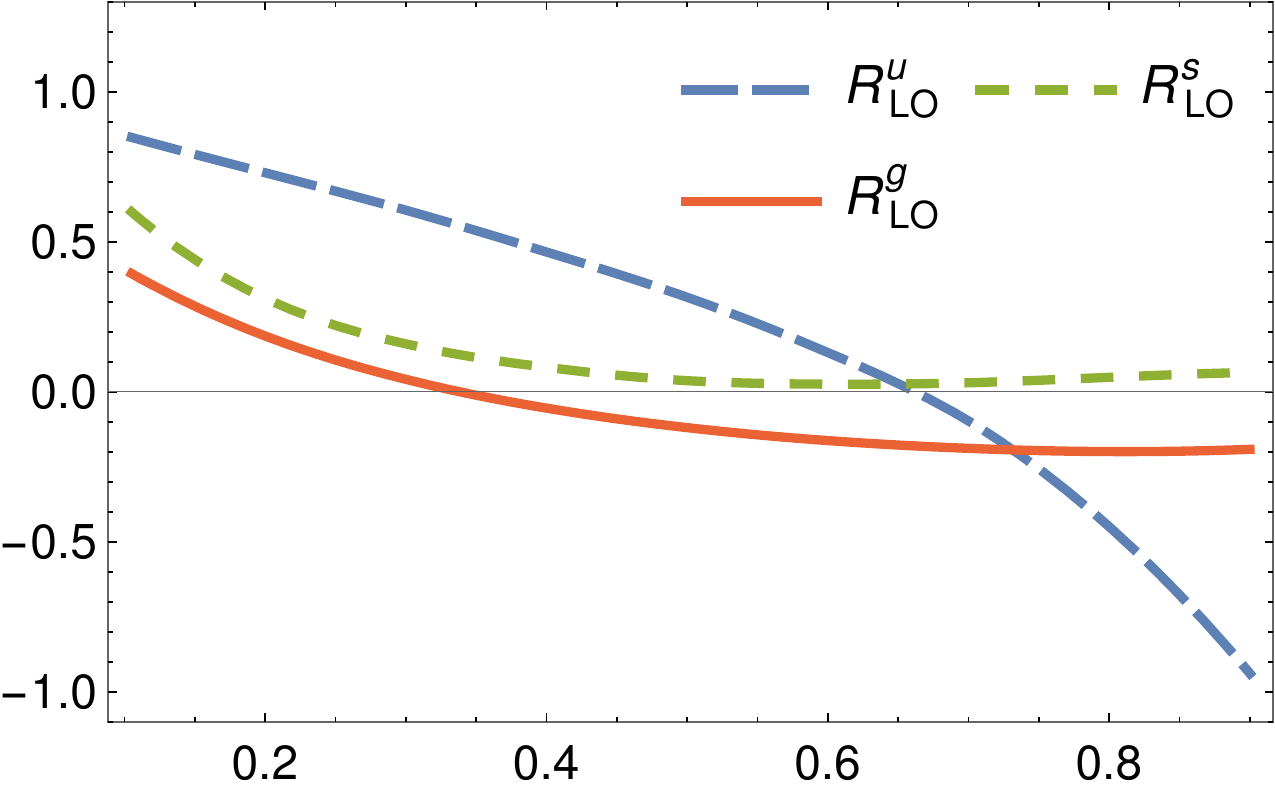}
          \put(55,-3){$\xi$}
    \end{overpic}
    \end{minipage}
    \vspace{4mm}
    \caption{Truncation error  of the leading order amplitude in the phenomenological model at $\mu^2=2$ GeV$^2$ (left) and  $\mu^2=10 \, \text{GeV}^2$ (right). }
    \label{xfit}
\end{figure}

\begin{figure}
    \centering
     \begin{minipage}{0.45\textwidth}
     \begin{overpic}[
         width=\textwidth]{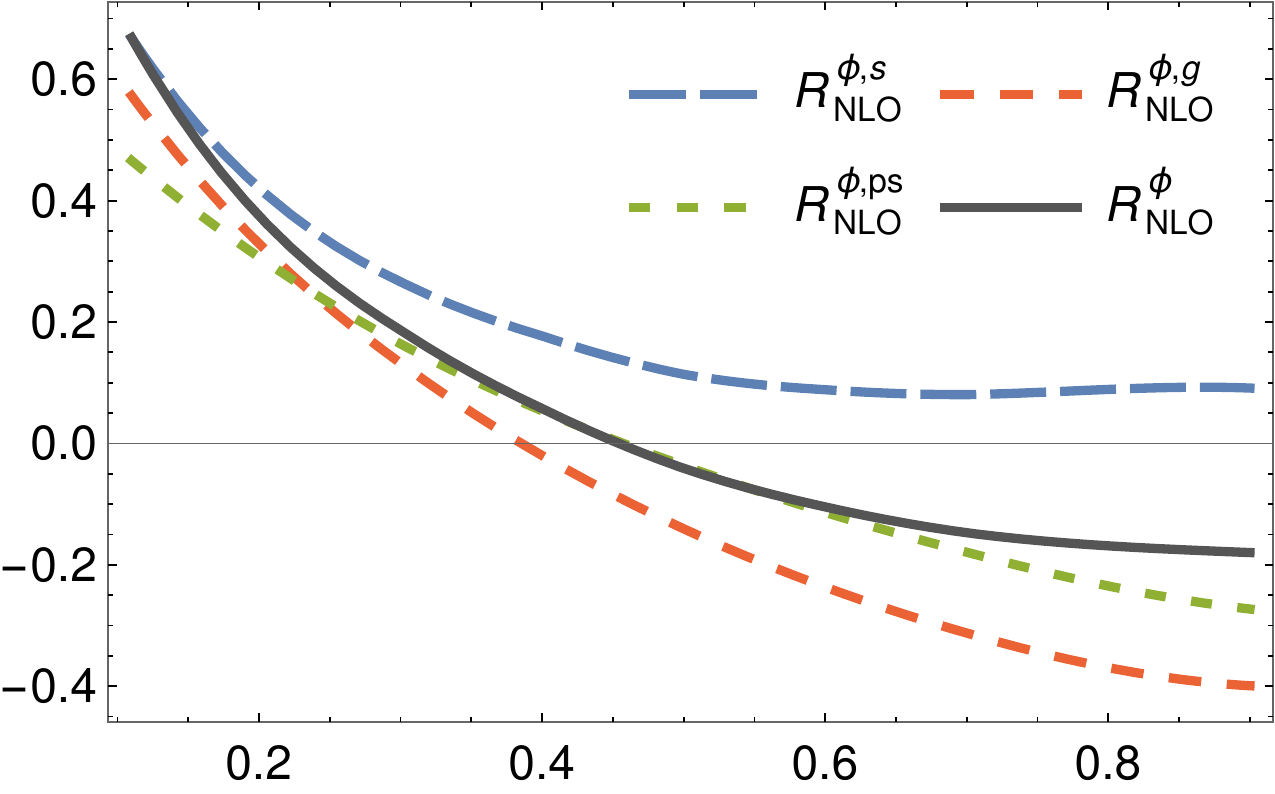}
          \put(55,-3){$\xi$}
     \end{overpic}
      \end{minipage}
      \hspace{0.05\textwidth}
    \begin{minipage}{0.45\textwidth}
    \begin{overpic}[
         width=\textwidth]{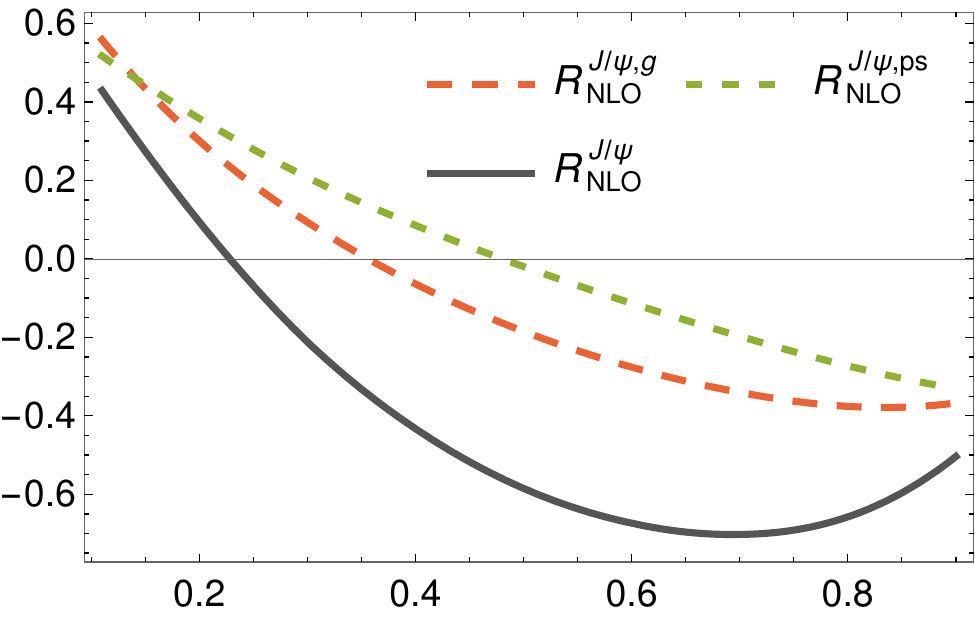}
          \put(55,-3){$\xi$}
    \end{overpic}
    \end{minipage}
    \vspace{4mm}
    \caption{Truncation error (\ref{error2}) of the full NLO amplitude in the phenomenological model. Left: $\phi$-electroproduction at $\mu^2=10$ GeV$^2$. Right:  $J/\psi$-photoproduction at $\mu^2=10$ GeV$^2$.  }
    \label{xfitfull}
\end{figure}

Fig.~\ref{xfit} shows the truncation error 
\beq
R^a(\xi) = 1-\frac{\sqrt{({\rm Re}\, {\cal H}^a)^2+({\rm Im}\, {\cal H}^a)^2}}{{\cal H}^a_{ \rm trunc}},
\label{error2}
\eeq
of the leading order amplitudes in the phenomenological model. At $\mu^2=2$ GeV$^2$ (left plot), the threshold approximation is good for strange quarks but marginally acceptable for gluons. 
This is due to the aforementioned slow growth of gluons at small-$x$, hence also in the central region $|x|<\xi$, making the  imaginary part $H^g(\xi,\xi)$ relatively non-negligible.  However, the situation improves at $\mu^2=10$ GeV$^2$ (right plot) since the evolution enhances the gluon density in the region $|x|<\xi$. Next, we plot the truncation error in the full NLO  amplitudes in Fig.~\ref{xfitfull}. 
The threshold approximation in  $\phi$-electroproduction (left plot) is acceptable, although not as good as for the algebraic model. On the other hand, the approximation fails in $J/\psi$-photoproduction (right plot). This is because of a large  cancellation between the gluon and quark terms in (\ref{nloJ}) which makes the denominator of (\ref{error2}) small. Although the quark terms only appear at NLO, it is numerically important because of the unusually small gluon momentum fraction $A_\pi^g(2\, {\rm GeV}^2)\approx 0.23$ of this model. A recent lattice QCD calculation \cite{Hackett:2023nkr} points to a much larger value $A_\pi^g(2\, {\rm GeV}^2)\approx 0.54$.  We thus expect that the failure of the phenomenological model does not generalize to more realistic models. (Note that  $A_\pi^g(2\, {\rm GeV}^2)\approx 0.41$  in the algebraic model.) Still, this example highlights  the significance of NLO corrections and illustrates  how the threshold approximation can fail due to an accidental cancellation.

\end{document}